\documentclass[useAMS,usenatbib]{mn2e}

\usepackage{graphicx,epsfig,amssymb,amsmath,layout,verbatim,rotating,calc,mathrsfs,natbib}
\usepackage{graphics}
\usepackage{epstopdf}
\usepackage{rotate}
\usepackage{txfonts}

\def\P{\bar{\Phi}}

\def\sles{\lower2pt\hbox{$\buildrel {\scriptstyle <}
   \over {\scriptstyle\sim}$}}
\def\sgreat{\lower2pt\hbox{$\buildrel {\scriptstyle >}
   \over {\scriptstyle\sim}$}}

\title[The pulsar synchrotron in 3D: curvature radiation]{The
pulsar synchrotron in 3D: curvature radiation}
\author[]{Ioannis Contopoulos
$^{1}$
\thanks{icontop@academyofathens.gr} and Constantinos Kalapotharakos
$^{1}$
\thanks{ckalapot@phys.uoa.gr}\\
$^{1}$ Research Center for Astronomy and Applied Mathematics,
Academy of Athens, Soranou Efessiou 4, Athens 11527, Greece}

\date{Accepted .......... Received ..........; in original form ..........}

\begin{document}

\maketitle

\label{firstpage}

\begin{abstract}
We investigate the strong electric current sheet that develops at
the tip of the pulsar closed line region through time dependent
three-dimensional numerical simulations of a rotating magnetic
dipole. We show that curvature radiation from relativistic
electrons and positrons in the current sheet may naturally account
for several features of the high-energy pulsar emission. We obtain
light curves and polarization profiles for the complete range of
magnetic field inclination angles and observer orientations, and
compare our results to recent observations from the Fermi
$\gamma$-ray telescope.
\end{abstract}


\section{Introduction}

Soon after the discovery of pulsars, people proposed that the most
natural process that may account for the origin of pulsar emission
is curvature radiation (Radhakrishnan \& Cooke~1969; Ruderman \&
Sutherland~1975). It has been suggested that electromagnetic
cascades above the surface of the neutron star form a dense
electron-positron plasma in which electrons and positrons move at
relativistic velocities along dipolar magnetic field lines
(Sturrock~1971; Daugherty \& Harding~1982).
Curvature radiation emitted by these particles escapes unimpeded
to infinity through this plasma only at frequencies well above the
frequency of plasma waves in the plasma frame of reference. In the
observer's frame of reference, this condition becomes $\nu >
\nu_{lim}$, where
\begin{equation}
\nu_{lim}\sim 2\sqrt{\gamma} \nu_p \equiv \left(\frac{4 e^2 \gamma
n}{\pi m_e}\right)^{1/2}\ .\label{nup}
\end{equation}
Here, $n$ is the total number density of electrons and positrons,
and $e$ and $m_e$ are the electron charge and mass respectively.
In that case, curvature radiation is polarized predominantly in
the plane of the particle trajectory. At frequencies below the
above limiting value, however, only the extraordinary wave escapes
freely from the pulsar magnetosphere (Barnard \& Arons~1986), with
the observed emission polarized perpendicularly to the magnetic
field line plane (Jackson~1975). Electromagnetic cascades
accelerate primary electrons to Lorentz factors $\gamma\sim
10^{6-7}$, and produce electron-positron pairs with Lorentz
factors $\gamma\sim 10^{2-3}$. Radio emission is supposed to
originate by these pairs at altitudes of about $10^{7-8}$~cm where
$\nu_{lim}\sim 20$~GHz (see Gil, Lyubarsky \& Melikidze~2004 and
references therein). High-energy emission is supposed to originate
at even higher altitudes by the extremely relativistic primaries.

In Contopoulos~2009 (hereafter C09), we proposed that the source
of the relativistic particles responsible for pulsar emission may
be quite different from that in the above canonical paradigm. We
have shown that, in the context of axisymmetric force-free ideal
relativistic magnetohydrodynamics, relativistic
positrons\footnote{We will hereafter refer to the sign of charge
and type of charge carriers that correspond to the aligned
rotator, that is to a rotator with an inclination angle between
the magnetic pole and rotation axis of less or equal to $90^o$.
Our results remain unchanged for a counter-aligned rotator,
provided we replace the sign of charge and type of charge carriers
by their opposites.} at the tip of the region of closed field
lines (hereafter the `dead zone') corotate with Lorentz factors
\begin{equation}
\gamma \sim \left(\frac{e B_* r_*^3 \Omega^2}{4 m_e
c^4}\right)^{1/2} \sim 10^4
\left(\frac{B_*}{10^{13}\mbox{G}}\right)^{1/2}
\left(\frac{P}{\mbox{sec}}\right)^{-1}\ , \label{gammaC09}
\end{equation}
and number densities
\begin{equation}
n \sim \frac{B_*^2 r_*^6 \gamma \Omega^6}{16 \pi m_e c^8} \sim
3\times 10^{12} \mbox{cm}^{-3}
\left(\frac{B_*}{10^{13}\mbox{G}}\right)^{5/2}
\left(\frac{P}{\mbox{sec}}\right)^{-7}\ .\label{nC09}
\end{equation}
Here, $B_*$ is the polar value of the stellar dipole magnetic
field; $\Omega$ is the stellar rotation angular velocity; and
$P\equiv 2\pi/\Omega$ is the stellar period of rotation. As they corotate,
these particles emit incoherent curvature radiation up to the
characteristic frequency
\begin{equation}
\nu_c \sim \frac{\gamma^3 c}{r_{lc}}\ ,\label{nuc}
\end{equation}
where, $r_{lc}\equiv c/\Omega_*$ is the radius of the so called
`light cylinder'. One can easily check that,
\begin{eqnarray}
\nu_c & \approx & \nu_{lim}\ \sim\ \left(\frac{e B_* r_*^3}{4 m_e
c^4}\right)^{3/2} \Omega^4 \nonumber \\
& \sim & 4\times 10^{12}
\left(\frac{B_*}{10^{13}\mbox{G}}\right)^{3/2}
\left(\frac{P}{33\mbox{ms}}\right)^{-4} \ .\label{nucnulim}
\end{eqnarray}
For Crab-like young pulsars, $\gamma\sim 10^5$ and $\nu_c\sim
10^{18}$~Hz (X-rays). For millisecond pulsars, $\gamma\sim 10^6$
and $\nu_c\sim 10^{22}$~Hz ($\gamma$-rays). C09 proposed that
radio emission is produced through some yet unknown coherence mechanism by the same particles that emit high-energy curvature radiation,
and must therefore, be in phase with the high-energy emission.  This
may indeed be the case in the Crab pulsar, but recent observations
from the Fermi $\gamma$-ray telescope suggest that this is not the
case in the vast majority of pulsars (we will address this issue
in \S~3 and 4). Notice that, although the scenario that pulsar
emission is coming from the light cylinder is not new (in the
early days of pulsar astronomy, people discussed the possibility
that pulses may be produced by hot plasma at discrete
positions on the light cylinder; e.g. Gold~1968, 1969;
Bartel~1978; Cordes~1981; Ferguson~1981), our picture is original
in several respects: the plasma is cold in its rest frame, the
fundamental synchrotron radiation frequency is the neutron star
rotation frequency, and pulses are due to the misalignment between
the rotation and magnetic axes.

Obtaining the structure of the pulsar magnetosphere turned out to
be a formidable problem that had to wait for about 40 years before
the first numerical solutions became available (Spitkovsky 2006;
Kalapotharakos \& Contopoulos~2009, hereafter KC09). Even the
axisymmetric problem described qualitatively in the seminal paper
of Goldreich \& Julian~(1969) had to wait for more than 30 years
for its resolution (Contopoulos, Kazanas \& Fendt~1999, hereafter
CKF; see also Uzdensky~2003; Gruzinov~2005; Contopoulos~2005;
Timokhin~2006). CKF generalized the discussion of Goldreich \&
Julian~(1969) and emphasized the importance, not only of the
space-charge density, but more so of the electric current. They
showed that, in order for an ideal MHD magnetosphere to fill all
space, it needs to be filled with a definite electric charge
density distribution $\rho_e$ (so called the `Goldreich-Julian
density', hereafter {\em the GJ density}) and a definite electric
current density distribution ${\bf J}$ (hereafter {\em the CKF
current}). A not so well appreciated fact is that the GJ electric
charge density depends not only on ${\bf \Omega} \cdot{\bf B}$ but
also on the CKF current. In the axisymmetric case, one can show
that
\begin{equation}
\rho_e = \frac{-{\bf \Omega} \cdot {\bf B} + \left(4\pi \Omega_* /
c^2\right) I (J_p/B_p)}{2\pi c (1-[r\sin\theta/r_{lc}]^2)}
\label{rhoGJ}
\end{equation}
where, $I$ is the electric current flowing through cylindrical
radius $r\sin\theta$; $J_p$ and $B_p$ are the poloidal
(meridional) electric current density and magnetic field
respectively; and $(r,\theta,\phi)$ are the usual spherical
coordinates centered on the star with the $\theta=0^o$ axis along
the axis of rotation.

Uzdensky~(2003) and Gruzinov~(2005) showed that the required GJ
charge at the tip of the dead zone diverges as the tip approaches
the light cylinder, and C09 showed that the dead zone will end at
a distance from the light cylinder where the inertia associated
with the corotating GJ charge carriers becomes comparable to the
magnetic stresses there. In other words, if the dead zone were to
extend even closer to the light cylinder, the magnetic field would
not have been able to hold the minimum amount of charge carriers
needed to supply the required local GJ density, and closed
magnetic field lines would be `ripped open' by centrifugal
relativistic inertial forces. Typical values for a Crab-like
pulsar are a distance from the light cylinder of only
$10^{-3}$~cm, a corotation Lorentz factor of about $10^5$, and a
typical positron/electron number density at the tip of the dead
zone on the order of $10^{23}$~cm$^{-3}$.

CKF showed that the CKF electric current circuit consists of three
parts: a) a distributed main magnetospheric electric current
flowing through the main part of the polar cap (the stellar
surface area around the two magnetic axes from which originate
open field lines), b) a distributed small amount of returning
electric current flowing through the rest of the polar caps, and
c) a singular main return current flowing along the separatrix
between open and closed field lines (hereafter the `separatrix')
that connects to the equator beyond the light cylinder. No
poloidal electric current flows inside the dead zone.
Interestingly enough, wherever $J_p$ becomes a surface/singular
current (that is along the separatrix and along the equator),
eq.~(\ref{rhoGJ}) yields a surface/singular charge density
\begin{equation}
\sigma_e = \frac{E^+ - E^-}{4\pi} \equiv
\frac{r\sin\theta}{r_{lc}}\frac{B_p^+ - B_p^-}{4\pi}\ ,
\label{sigma}
\end{equation}
where ${\bf E}\equiv (r\sin\theta/r_{lc}){\bf B}_p\times
\hat{\phi}$ is the electric field; $(\ldots)^\pm$ are the field
components outside/inside the separatrix or above/below the
equator respectively.
We find that the separatrix
is charged with a negative surface charge, whereas the equator
with a positive one. We emphasize once again that these surface
charge and current densities are required by the condition of
ideal MHD in the magnetosphere, therefore, they too are part of
the GJ charge and the CKF current.

In the present paper, we would like to extend the earlier
axisymmetric investigation in three dimensions (hereafter 3D), as
required by the fundamentally 3D nature of the pulsar phenomenon.
Our goal is to use the GJ charge and CKF current distributions
obtained in the numerical simulations of KC09 to derive curvature
radiation emission and polarization profiles for all inclination
and observation angles. In \S~2, we will present the main findings
of our numerical work, with particular emphasis on the strong
electric current sheet that develops at the tip of the pulsar
closed line region. In \S~3, we will derive curvature radiation
pulse light curves and polarization profiles for various values of
the inclination and observation angles. We will end in \S~4 with a
summary of our results.

\section{Numerical simulations}

\begin{figure*}
\begin{center}
\includegraphics[width=13.2cm]{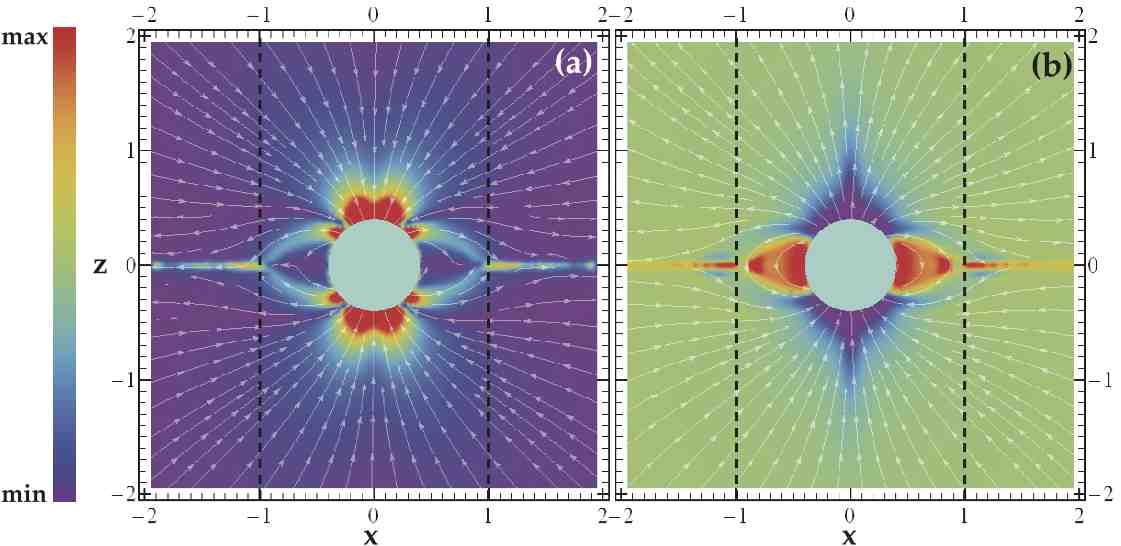}
\end{center}
\caption{Meridional cross section of the magnetosphere of an
aligned (axisymmetric) rotator ($\alpha=0^o$). Left panels:
meridional electric current density (color plot) and direction
(arrows). Right panels: electric charge density (color plot) and
meridional magnetic field direction (arrows). Light cylinder shown
with dashed lines.} \label{fig1}
\end{figure*}
\begin{figure*}
\begin{center}
\includegraphics[width=13.2cm]{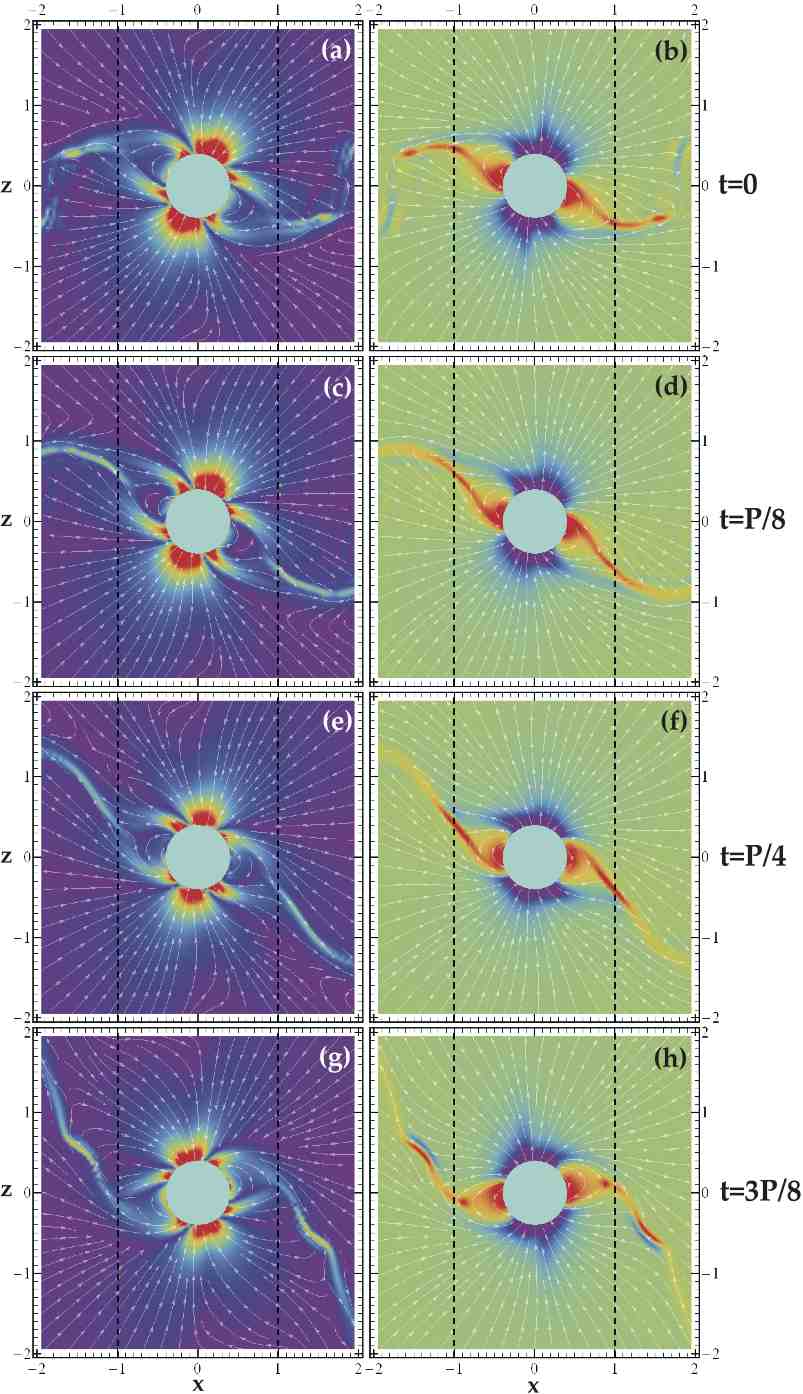}
\end{center}
\caption{Same as Fig.~1 only for an inclined rotator with
$\alpha=30^o$ at times $t=0, P/8, P/4, 3P/8$. At $t=0$, the
magnetic axis lies in the $(x,z)$ plane.} \label{fig2}
\end{figure*}
\begin{figure*}
\begin{center}
\includegraphics[width=13.2cm]{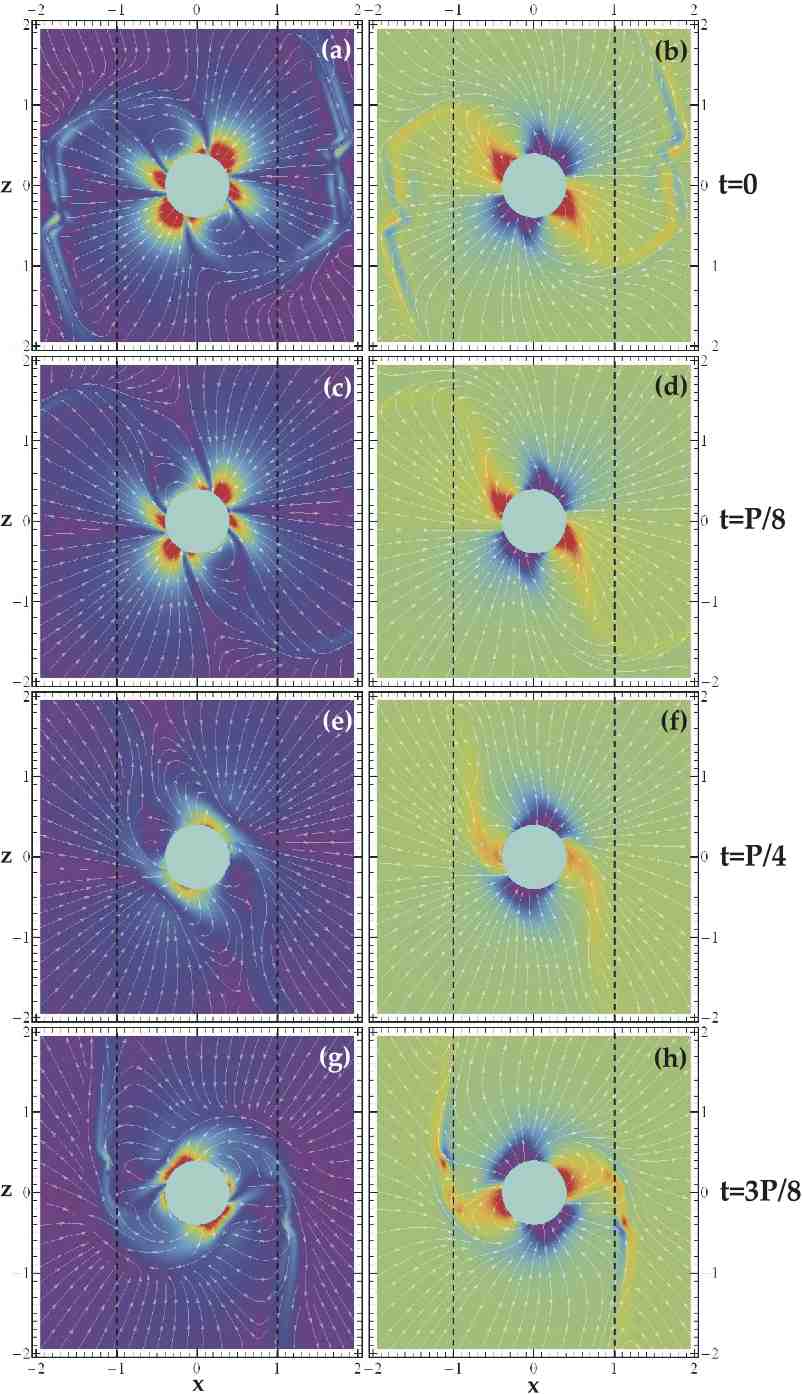}
\end{center}
\caption{Same as Fig.~2 for $\alpha=60^o$.} \label{fig3}
\end{figure*}
\begin{figure*}
\begin{center}
\includegraphics[width=13.2cm]{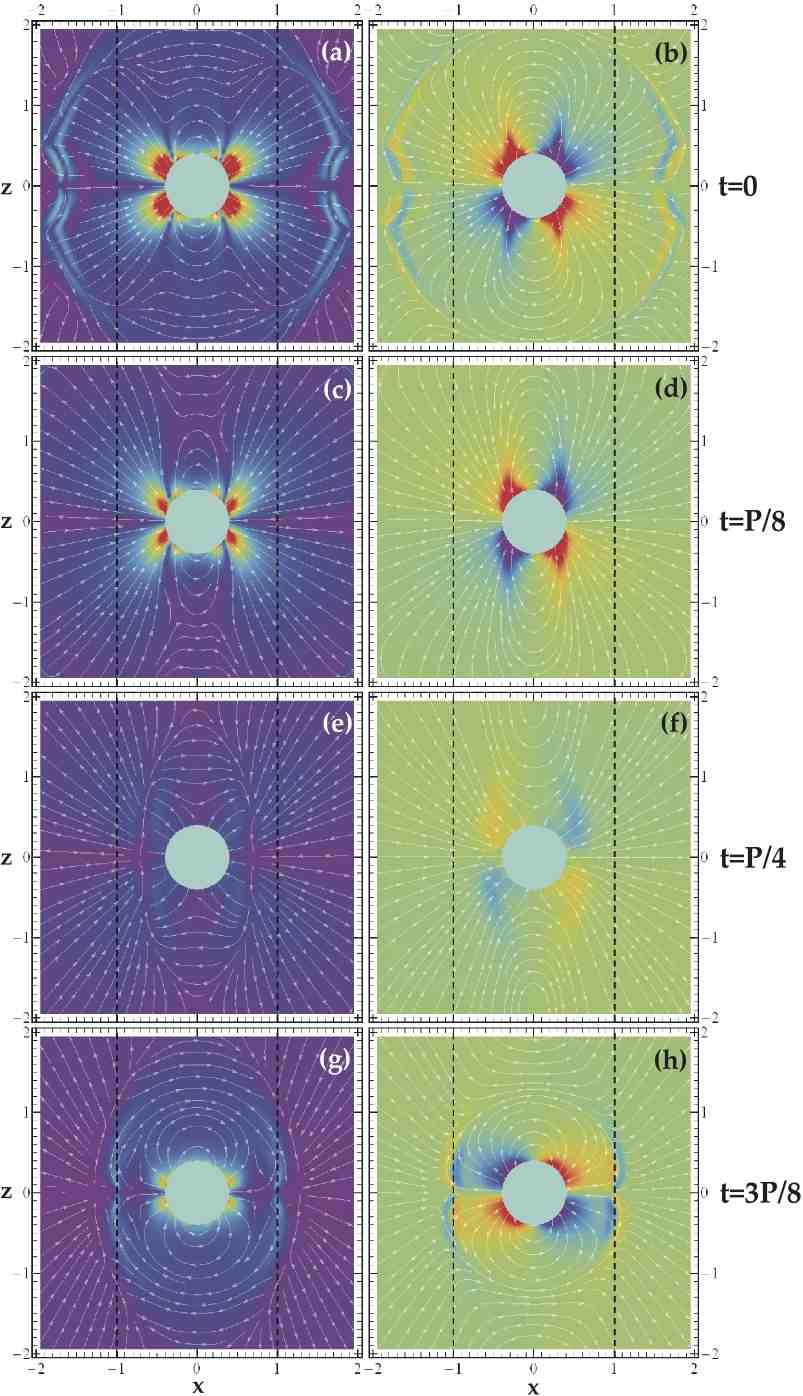}
\end{center}
\caption{Same as Fig.~2 for $\alpha=90^o$.} \label{fig4}
\end{figure*}

We have assumed in the past that force-free relativistic ideal MHD
is a valid description of the pulsar magnetosphere, except for
boundary regions of infinitesimal volume where ideal MHD breaks
down and dissipation takes place.\footnote{We imply here that
electric charges are `freely' available in the magnetosphere
wherever they are needed to supply the required GJ density and CKF
current. Most researchers assume that this takes place in
so-called `gaps' where ideal MHD breaks down locally, particles
are accelerated to extremely relativistic velocities by electric
fields parallel to the magnetic field, and electron-positron pairs
are copiously produced to supply the required GJ charges and CKF
currents. But not everybody agrees with that statement. An
alternative scenario where charges are not freely supplied
wherever they are needed by ideal MHD is strongly voiced in
Michel~2005.} Therefore, the magnetospheric electric and magnetic
fields ${\bf E}$ and ${\bf B}$ satisfy the three time-dependent
Maxwell's equations in the presence of electric charges
$\rho_e\equiv (4\pi)^{-1}\nabla\cdot {\bf E}$ and electric
currents ${\bf J}$,
\begin{equation}
\frac{\partial {\bf E}}{\partial t} = c \nabla\times {\bf B} -
4\pi {\bf J} \ , \label{M1}
\end{equation}
\begin{equation}
\frac{\partial {\bf B}}{\partial t} = -c \nabla\times {\bf E}\ ,
\end{equation}
\begin{equation}
\nabla\cdot {\bf B}=0\ ,
\end{equation}
as well as the ideal MHD condition
\begin{equation}
{\bf E}\cdot {\bf B} = 0 \ . \label{M4}
\end{equation}
As is, the system of equations is degenerate since we need one
extra equation for ${\bf J}$. This is provided through the
force-free condition
\begin{equation}
\rho_e {\bf E} + {\bf J}\times {\bf B}=0\ , \label{forcefree}
\end{equation}
which, together with eqs.~(\ref{M1})-(\ref{M4}), allows us to
express ${\bf J}$ as a function of ${\bf E}$ and ${\bf B}$, namely
\begin{equation}
{\bf J} = \frac{c}{4\pi}\nabla \cdot {\bf E}\ \frac{{\bf E}\times
{\bf B}}{B^2} + \frac{c}{4\pi}\frac{({\bf B}\cdot \nabla\times
{\bf B} - {\bf E}\cdot \nabla\times {\bf E})}{B^2}\ {\bf B}
\label{J}
\end{equation}
(Gruzinov 1999). The first term in eq.~(\ref{J}) corresponds to
the Goldreich-Julian charge density moving at the magnetospheric
drift velocity $c{\bf E}\times {\bf B}/B^2$, whereas the second
one corresponds to the extra field-parallel electric current
required by the ideal MHD condition that ${\bf E}$ and ${\bf B}$
are everywhere orthogonal (eq.~\ref{M4}). Note that, whenever
current/charge sheets appear, they are treated as unresolved MHD
discontinuities that satisfy continuity of the Lorentz invariant
${\bf E}^2-{\bf B}^2$ (our ideal MHD formulation does not capture
the non-ideal MHD physical conditions inside a current/charge
sheet).

As described in KC09, we start with an initially stationary
dipolar magnetic field configuration at a certain inclination
angle $\alpha$ (that configuration obviously satisfies the set of
equations \ref{M1}-\ref{forcefree}), and we then set the central
star in rotation. After several stellar rotations (1.5 to 4
depending on the grid numerical diffusivity) the magnetosphere
relaxes to a stationary corotating pattern. We produced our own
Eulerian
solver based on the Yee algorithm (Yee~1966) which we run on a
standard PC with 2~Gbyte of RAM. Our
computational grid consists of a cubic region with sides 5 times
$r_{lc}$ centered around the star, with grid resolution
$\delta=0.04 r_{lc}$ (we also run a few cases with $\delta=0.025
r_{lc}$; see Appendix). In order to reduce computer memory
requirements we made use of the central symmetry of the problem
and thus integrated only on one half of the above cube.
Furthermore, in order to be able to follow the evolution of the
magnetosphere for several stellar rotations, we implemented the
technique of Perfectly Matched Layers (PML) for our outer
boundaries (Berenger~1994, 1996). This technique guarantees
non-reflecting and absorbing boundary conditions in vacuum. By
applying it to our problem, we showed in practice that this
technique also works in non-vacuum relativistic ideal force-free
MHD.

Our steady-state results may be summarized in the time sequences
of various physical quantities shown on the rotational meridional
plane in Figs.~(1)-(4) (times in units of one stellar rotation
period $P$, with $t=0$ defined as the instance of time when the
magnetic axis lies on the plane shown). These are obtained for
various inclination angles $\alpha$ after steady state is reached
and a stationary corotating electromagnetic field pattern is
safely established. Due to the obvious central symmetry of the
problem, we plot the time evolution only through one half-period
since the pattern repeats itself. On the left panels we plot the
distribution of meridional electric current density (the arrows
show its direction, and the color plot its magnitude) on a
meridional cross section of the magnetosphere. On the right panels
we plot the distribution of meridional magnetic field and electric
charge density. The toroidal electric current induced by the
corotation of the electric charge distributed throughout the dead
zone is not reflected in the color diagrams of the left panels.

In the axisymmetric case we obtained a steady-state configuration
after about 2.5 stellar rotations (Fig.~1). We recovered the
following important characteristics predicted in CKF and others:
\begin{enumerate}
\item The dead zone grows up to the light cylinder. Its growth is
limited by random reconnection events that occur when its tip
accidently crosses the light cylinder. This results in the random
generation of outward moving equatorial plasmoids (e.g.
Bucciantini {\em et al.}~2006). \item The main magnetospheric
poloidal current circuit closes only partially through the two
magnetic polar caps. The main part of the return current flows
along both surfaces of the separatrix (upper and lower), and
connects with an equatorial current sheet at the tip of the dead
zone. Interestingly enough, the two surfaces of the separatrix are
negatively charged, whereas the equator is positively charged (see
eq.~\ref{sigma}). \item The space charge density grows at the tip
of the dead zone giving rise to an azimuthal electric current
density much stronger than that in the return electric current
sheet (C09).
\end{enumerate}
An interesting feature of our numerical method is that the
equatorial current sheet is captured well (it is 1-2 computational
cells thick) because it coincides with one of our cartesian grid
directions. The separatrix current sheet is, however, wider (it is
3-4 computational cells thick). Notice a slight limitation of our
numerical scheme: magnetic field lines at high latitudes
($\sgreat\ 65^o$) reach the boundaries of our computational box
before they cross the light cylinder. As a result, the
distribution of GJ charge density and CKF electric current are not
captured accurately near the rotation axis (see artificial
helmet-like enhancements in Fig.~1b; also in Figs.~2b, 3b).
Fortunately, the overall effect of that region in the global
magnetospheric structure is only minimal.

This simple picture evolves as we move to non-zero inclination
angles $\alpha$. In that case we obtained steady-state
configurations after about 1.5 stellar rotations (Figs.~2-4). We
observed the following characteristic features:
\begin{enumerate}
\item The dead zone again grows up to the light cylinder. Random
plasmoid formation is again going on beyond the light cylinder
(e.g. small knots just outside the light cylinder in Fig.~2g), but
is not captured (numerically) as well as in the axisymmetric case
(we do see plasmoid formation in the higher resolution simulations
shown in Fig.~\ref{resolution} though). \item As the inclination
angle $\alpha$ approaches $90^o$, a larger and larger amount of
the return current is distributed across the polar cap. An
inclination angle of $90^o$ is the limit of both an aligned and a
counter aligned rotator, and therefore, at $90^o$ the two signs of
GJ charge and CKF current are expected to be distributed equally
across both polar caps, as is seen. \item A well resolved
(numerically) ondulating equatorial current sheet originates at
the tip of the dead zone on the light cylinder. This is true even
in the orthogonal rotator ($\alpha=90^o$). Obviously, the electric
circuit cannot close through the dead zone because, in ideal MHD,
closed magnetic field lines do not support out/in-flowing electric
currents. Therefore, this singular electric current can only be
supplied in the form of a current sheet along the separatrix
between open and closed field lines. As in the axisymmetric case,
electric current sheets inside the light cylinder are not captured
well by the numerical simulation (we do resolve parts of the
current sheets in Figs.~2c, 2e, 3e, 3g, \ref{resolution} though).
Nevertheless, the observation of strong well resolved (1-2
computational cells wide) electric charge sheets inside the light
cylinder (e.g. Figs.~2b, 2d, 2f) and eq.~(\ref{rhoGJ}) (i.e. the
one-to-one relation between charge and current sheets) further
support the idea that the singular return currents persist, at
least over some part of the separatrix.\footnote{In the finishing
stages of our work we became aware of a new paper by Bai \&
Spitkovsky~2009 that presents a different opinion on this issue.}
Interestingly enough, when axisymmetry is broken ($\alpha
> 0^o$), the leading part of the separatrix (that is the surface
that faces in the direction of rotation) remains charged
negatively, whereas the trailing part is charged positively (we
checked that this indeed the case for $\alpha \geq 15^o$). On the
other hand, the equatorial current sheet remains always charged
positively on average. \item The enhancement of the space charge
density at the tip of the dead zone is present in the
non-axisymmetric case too, at least over part of the tip (see
Figs.~2h, 3h). By analogy to the axisymmetric case, this results
in a strong electric current in the return current sheet, both in
the azimuthal and outward direction. \item A novel interesting
feature of the numerical solution is that at an azimuthal angle of
about $90^o$ to that of the magnetic axis, originates a strong
source of positive separatrix surface charge
(in Fig.~2 notice the strong singular structure that seems to grow
on the light cylinder at times $t=P/4$ and is seen moving outward
through time $t=P/2$, or equivalently $t=0$). This outflowing
positive charge is surrounded by negative charges that guarantee
global electric charge conservation. Obviously, any structure
observed to move outward in a meridional cross section of the
magnetosphere corresponds to a corotating spiral in 3D. We are
well aware that a few of the features that we observe in our
simulations are due to our numerical resolution (we tested that by
running simulations at different resolutions, and found that the
details vary slightly; see the Appendix). We observed that this
spiral structure survives over several stellar rotations, and
therefore, we confirmed that it is steady. In fact, as we will see
in the next section, this structure may also leave a signature in
the observed pulsar emission.
\end{enumerate}

\section{Curvature radiation}

As we argued in C09 for an axisymmetric rotator, the tip of the
dead zone may play a key role in the pulsar phenomenon.
It is impossible to investigate this region in detail in a global
magnetospheric numerical simulation.
We thus have to rely on an extrapolation of the results in C09
guided by the low resolution numerical simulations of KC09. We
believe that, as in the axisymmetric case, the electric charge and
electric current density both diverge as the tip of the dead zone
approaches the light cylinder. Moreover, the dead zone must end at
a short distance from the light cylinder where equipartition is
reached between the particle inertia and the magnetic field. At
that position, the magnetosphere can only support one type of
charge carriers, namely {\em those and only those} needed to
supply the required GJ density. Therefore, {\em in the current
sheet at the tip of the dead zone and beyond}, electric charges
will move along the direction of the local CKF current ${\bf J}$
with velocities
\begin{equation}
{\bf v} = \frac{\bf J}{\rho_e}\ . \label{speed}
\end{equation}
In analogy to C09, we expect these to correspond to Lorentz
factors $10^{5-6}$. The particle trajectories in the current sheet
are obviously not straight (at least near the light cylinder),
thus the particles will emit curvature radiation along their
direction of motion up to a cutoff frequency determined by
eq.~(\ref{nuc}). We note that, in the axisymmetric case where we
know analytically the distribution of electric and magnetic fields
inside and around the separatrix and the equatorial current sheets
(Uzdensky~2003; C09), we are able to calculate in detail the
meandering orbits of the particles that support the singular CKF
current. These calculations yield a whole spectrum of Lorentz
factor values. We therefore expect that the resulting spectrum of
curvature radiation is not that of a mono-energetic particle. We
plan to investigate this in a forthcoming publication where we
will also consider the effect of inverse Compton scattering on
soft photons coming from the stellar surface.

We are now in a position to determine the pattern of emitted
curvature radiation, both its light curve and polarization
profile, for various pulsar inclination and observation angles.
The local radiation intensity is weighted by a) the
Goldreich-Julian number density, b) the inverse square of the
particle instantaneous radius of curvature, and c) the fourth
power of the particle Lorentz factor $\gamma$ (for acceleration
perpendicular to the direction of motion; Jackson~1975). The
simulation provides the distribution of electric charge density
$\rho_e$, and the characteristic elements of the particle orbits
in the observer's inertial frame through eq.~(\ref{speed})
(direction, radius of curvature), but is not good enough to yield
accurately the Lorentz factor $\gamma$. We will thus assume simply
that positrons/electrons attain a very large limiting Lorentz
factor ($\gamma\sim 10^6$) at the tip of the separatrix and in the
equatorial current sheet. In practice, we implement the following
numerical procedure:
\begin{figure*}
\begin{center}
\includegraphics[width=12cm]{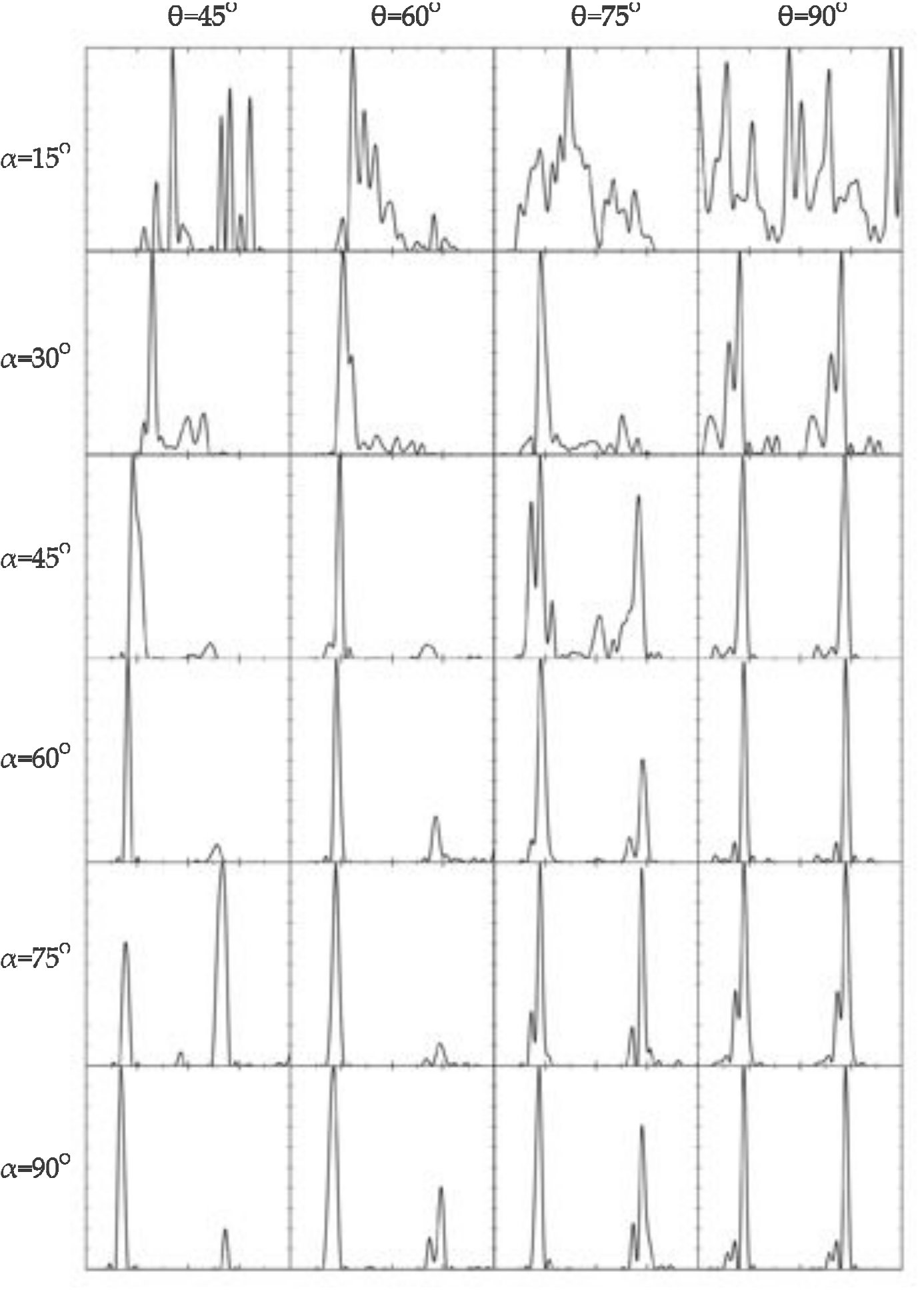}
\end{center}
\caption{High-energy light curves obtained for various inclination
and observation angles $\alpha$ and $\theta$. Horizontal axes are
normalized to the period of one stellar rotation, with zeros
corresponding to the arrival of the radio pulse (see text).
Vertical axes are normalized to the maximum pulse intensity.}
\label{fig5}
\end{figure*}
\begin{enumerate}
\renewcommand{\theenumi}{(\arabic{enumi})}
\item We identify the regions of the magnetosphere where the
particle velocities $\bf v$ (as given by eq.~\ref{speed}) are
larger than $0.9c$. We also require that the particle density
$\left|\rho_e\right|$ be above a certain value in order to avoid
regions with fictitiously high values of $\bf v$. This selection
indeed coincides with the ondulating equatorial current sheet
outside the light cylinder. \item We assume that every point in
the selected region emits in a forward small opening angle
($2^o-5^o$) around its instantaneous direction of
motion.\footnote{The relativistic particle trajectories are
calculated in the observer's inertial frame through
eq.~(\ref{speed}), and therefore, the calculation of the direction
of emission is much simpler than if we had obtained the particle
trajectories in the pulsar's rotating frame (as for example in
Harding {\em et al.}~2008).} \item The local curvature radiation
intensity is weighted proportionally to the local particle density
$|\rho_e|$ and inversely proportionally to the square of the local
radius of curvature of particle orbits. There is no weighting with
$\gamma^4$ since, as we said, $\gamma$ is set to the same large
value in the equatorial current sheet. \item We select the
direction of observation and add up the contributions along that
direction from every point in the region selected in (1) and (2)
weighted as in (3) during one full pulsar rotation. In doing so,
we take into account the light-crossing time-delay from the
various parts of the magnetosphere.
\end{enumerate}

\begin{figure}
\begin{center}
\includegraphics[width=8cm]{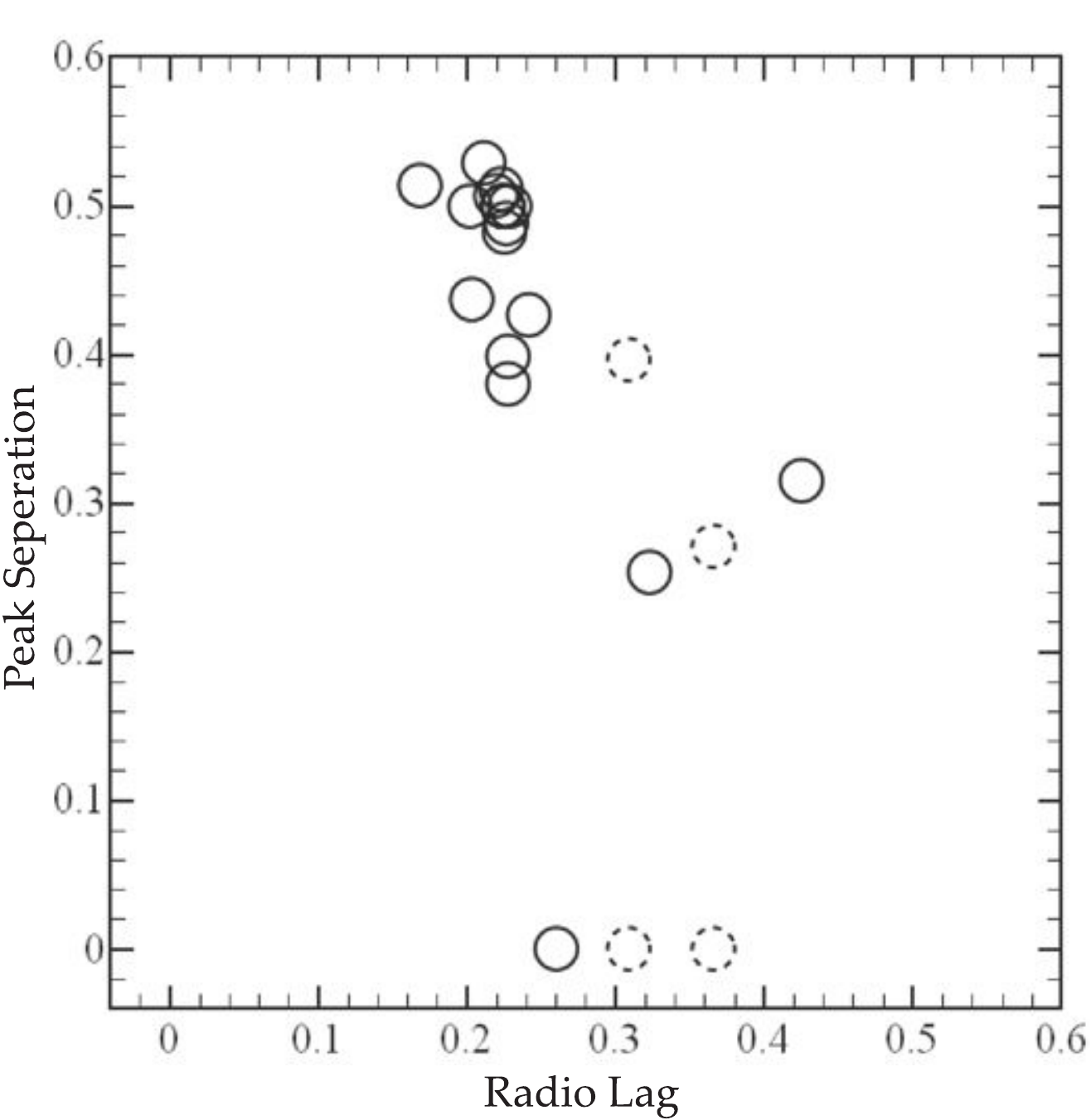}
\end{center}
\caption{Phase difference between the high-energy peaks, versus
phase lag between the radio and main high-energy peak. Dotted
circles correspond to those cases where interpulses are not
clearly defined.} \label{lag-sep}
\end{figure}

The light curves produced by curvature radiation in the equatorial
current sheet from the tip of the dead zone on the light cylinder
and beyond are shown in Fig.~(\ref{fig5}). We plot pulse profiles
with intensities normalized to the maximum pulse intensity for
various inclination and observation angles $\alpha$ and $\theta$
(calculated with respect to the axis of rotation). In the context
of our model of equatorial current sheet emission, emission
intensity dramatically decreases at high latitudes, and the
quality of our numerical results deteriorates. This is why we only
plot light curves for $\theta \geq 45^o$. Azimuthal phase zero is
defined as the phase where a photon that is emitted radially
outward from the surface of the star when the magnetic axis lies
in the plane defined by the rotation axis and the line of sight
reaches the observer. As defined, azimuthal phase zero corresponds
to the arrival phase of the radio pulse in the prevailing model of
pulsar radio emission coming from a magnetospheric region just
above the magnetic polar cap. Several interesting features in the
pulse/sub-pulse distribution are worth mentioning:
\begin{enumerate}
\item Pulses are narrower when observed from higher latitudes.
\item Pulses are narrow despite the fact that the corresponding
emission regions have significant azimuthal extent. This effect is
analogous to the `caustics' obtained in pulsar slot-gap models for
radiation emitted along trailing magnetic field lines (e.g. Arons
1983; Harding {\em et al.}~2008).\footnote{A similar effect is
described in Bai \& Spitkovsky~2009.} \item Interpulse intensity
decreases fast compared to that of the main pulse as the observer
moves away from the rotational equatorial plane. \item Whenever an
interpulse is seen ($\alpha$ greater than about $30^o$),
pulse-interpulse separation varies mostly between about 0.4 to 0.5
times the period (Fig.~\ref{lag-sep}). The closer the inclination
or the observation angle is to $90^o$, the closer the
pulse-interpulse separation is to one-half of the period. \item
The main pulse trails the radio pulse (in the prevailing model of
radio pulsar emission coming from above the stellar polar cap) by
about 0.15 to 0.25 times the period whenever an interpulse is
seen, and up to one-half of the period when no interpulse is seen.
In particular, notice the striking similarity between our
Fig.~(\ref{lag-sep}) and Fig.~4 of Abdo {\em et al.}~(2009).
\end{enumerate}

We would like to end this section with an investigation of the
expected polarization angle profiles. Radiation propagates in the
magnetosphere in the form of ordinary and extraordinary wave modes
(Barnard \& Arons 1986). The ordinary mode is polarized in the
plane of the wave vector and the local magnetic field direction,
whereas the extraordinary mode is linearly polarized
perpendicularly to the wave vector and the local magnetic field.
Gil, Lyubarsky \& Melikidze~(2004) showed that,
for radiation frequencies $\nu\leq
\nu_{lim}$, the ordinary mode is heavily suppressed, and only the
extraordinary mode escapes freely and thus reaches the observer.
In our case $\nu_c\approx \nu_{lim}$ in the emission region
(eq.~\ref{nucnulim}), and therefore, we will heretofore consider
only the radiation mode linearly polarized perpendicularly to the
line of sight and the local magnetic field.

\begin{figure*}
\begin{center}
\includegraphics[width=12cm]{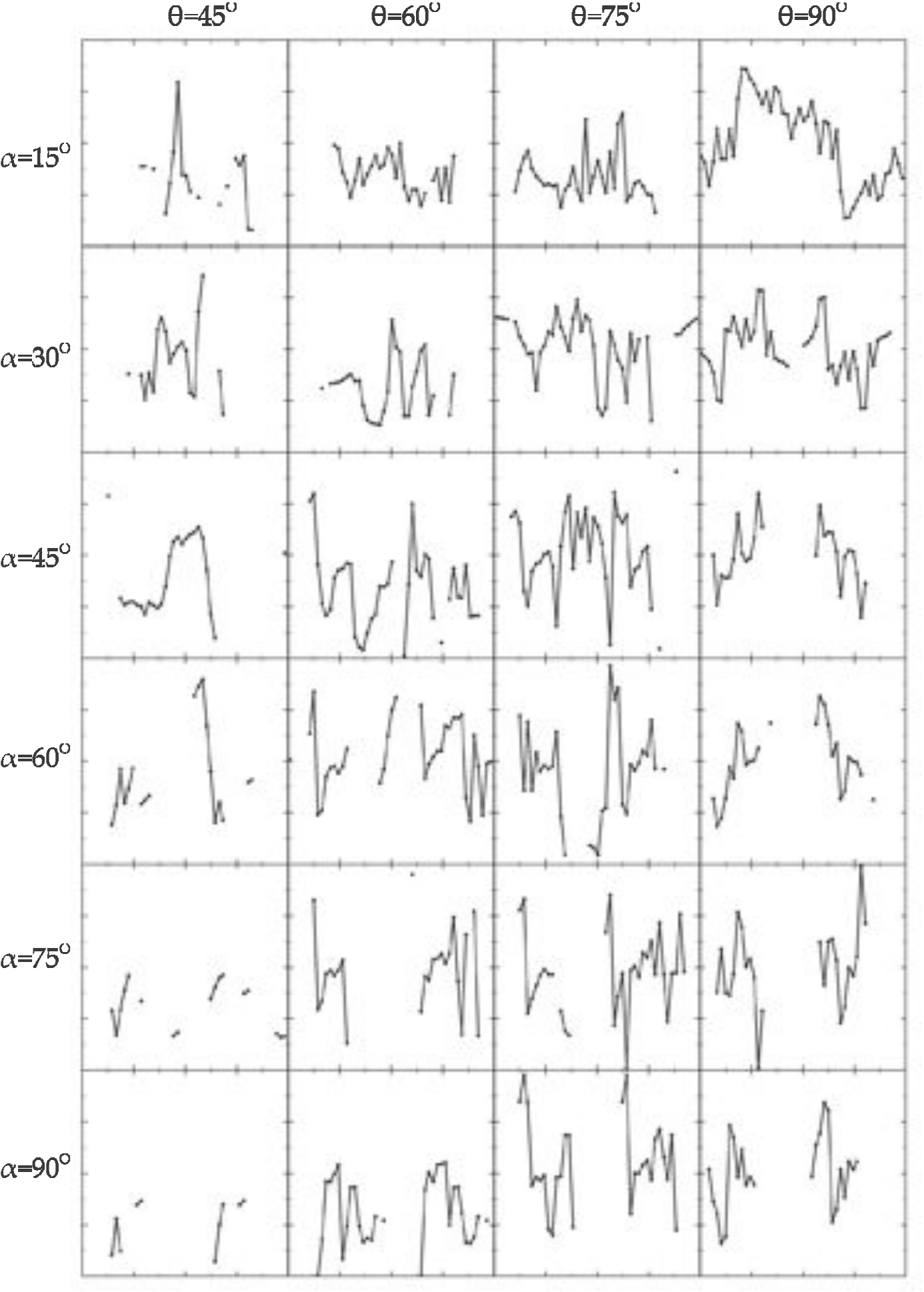}
\end{center}
\caption{High-energy polarization angle profiles obtained for
various inclination angles $\alpha$ and various observation angles
$\theta$. Horizontal scale as in Fig.~5. Vertical scale from
$-90^o$ to $+90^o$.} \label{fig9}
\end{figure*}

\begin{figure}
\begin{center}
\includegraphics[width=8cm]{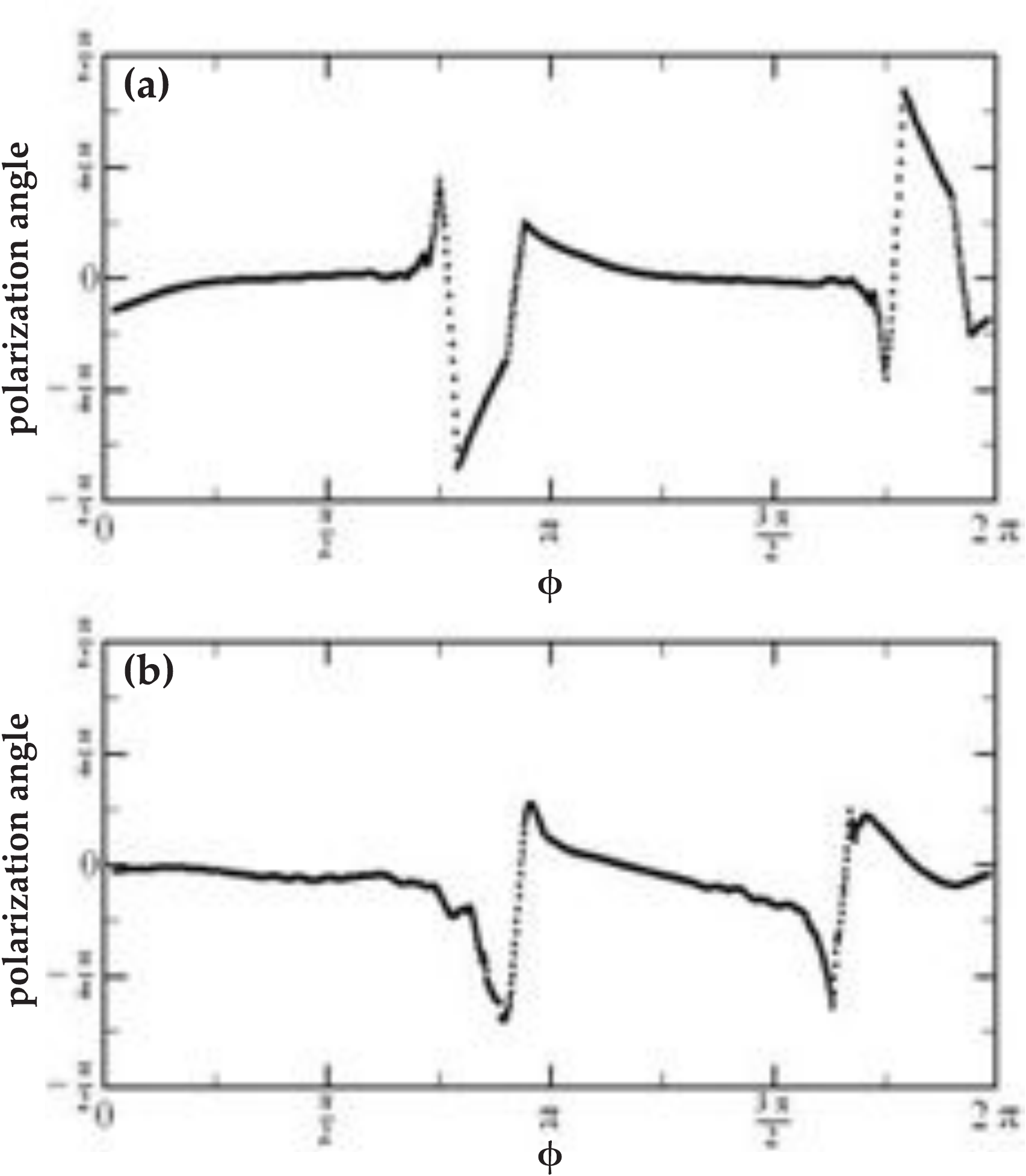}
\end{center}
\caption{Expected polarization angles from curvature radiation
produced in two annular regions at positions $r \sin\theta=1.5
r_{lc}$, $r\cos\theta = 0$ and $0.5 r_{lc}$ (a and b respectively)
as a function of azimuthal angle $\phi$ for an inclined rotator
with $\alpha=50^o$.} \label{rot-vec}
\end{figure}

In Fig.~(\ref{fig9}) we show the polarization profiles that
correspond to the light curves in Fig.~(\ref{fig5}). The
polarization angles shown are the mean polarization angles from
all the emission points observed at each time instance weighted by
their relative intensity. They lie in the range $-90^o$ to $+90^o$
and are measured on the plane of the sky with zero along the
projected direction of the axis of rotation, and positive in the
north-to-east direction. Our results are not very clear because of
numerical problems associated with the equatorial current sheet.
We do observe several cases with dramatic (up to $\pm 90^o$)
polarization angle sweeps across the main pulses and interpulses.
In particular, at observation angles $\theta=60^o$ and $75^o$, we
obtain polarization angle sweeps reminiscent of those seen in the
Crab pulsar (e.g. S\l owikiwska {\em et al.}~2009). In order to
explore the origin of this effect, we plot in Fig.~(\ref{rot-vec})
the polarization angles expected from radiation produced in $2\pi$
annular regions around the rotation axis if one happens to be
looking along the direction of the local ${\bf J}$. The observed
polarization angles sweeps are associated with the abrupt change
of magnetic field direction at the crossings of the equatorial
current sheet. We thus conclude that, in the context of our model,
polarization angle sweeps essentially cut across and resolve the
thin equatorial current sheet, as in the phenomenological model of
Petri \& Kirk~2005.

\section{Summary}

2D and 3D numerical simulations of the force-free relativistic
pulsar magnetosphere show that a strong electric current develops
at the tip of the closed line region on the light cylinder. This
electric current consists of extremely relativistic electrons and
positrons that move along curved trajectories in the azimuthal and
outward direction in an equatorial current sheet beyond the light
cylinder. As in C09, we call this configuration {\em the pulsar
synchrotron}.

In the present paper, we studied the emission of curvature
radiation from the pulsar synchrotron, and defered the study of
inverse Compton to a future publication. We showed that curvature
radiation seems to be responsible for the high-energy pulsed
emission up to X-ray frequencies in normal pulsars, and up to
$\gamma$-ray frequencies in millisecond pulsars. We produced
high-energy light curves that show sharp pulses and interpulses in
pulsars with high inclination angles $\alpha\ \sgreat\ 45^o$, and
wide single pulses in pulsars with low inclination angles $\alpha\
\sles\ 30^o$ (except when observed from very low latitudes with
respect to the rotation equatorial plane). Our model of an
extended emission region in the equatorial direction is in some
sense complimentary to that of the polar cap `lighthouse beam'
emission. Polarization angle profiles are less well determined
numerically. They do show, however, abrupt polarization angle
sweeps that are associated with the abrupt change of direction of
the magnetic field as we cut across the equatorial current sheet.

Our conclusions are based on our preliminary numerical simulations
of the 3D force-free relativistic pulsar magnetosphere. In
particular, we focus our investigation on the most difficult part
of our simulation, the return-current sheet. This may explain
certain strange features in our compilation of light curves (e.g.
for $\theta=45^o$, $\alpha=15^o$ and $75^o$), and the abrupt
changes in the polarization angle profiles. On the other hand,
some of these features may be attributed to the newly discovered
equatorial spiral structure that originates at $90^o$ from the
magnetic axes on the tip of the closed-line region on the light
cylinder. A deeper investigation of the equatorial return-current
sheet will have to wait our forthcoming supercomputer numerical
simulations (expected improvement in the numerical resolution by a
factor of 5 to 10).

Regarding the expected lag between the radio and high-energy
pulses, our results compare well with newly available data from
NASA's Fermi $\gamma$-ray space telescope (Abdo {\em et
al.}~2009). This comparison assumes that radio emission is coming
from a small height above the stellar magnetic polar cap.
Interestingly enough, radio emission from Vela is on average
polarized in a direction perpendicular to the axis of rotation
(assumed to coincide with the elongation axis of the nebula). This
is consistent with the observed pulsar radio emission consisting
mainly of extraordinary waves produced by relativistic electrons
and positrons that move along open dipolar magnetic field lines
above the polar cap (e.g. Lai, Chernoff \& Cordes~2001; Gil,
Lyubarsky \& Melikidze~2004). On the other hand, high-energy
(optical) emission from Vela is on average polarized along the
rotation axis as predicted by our model (Mignani {\em et
al.}~2007; Fig.~\ref{rot-vec}). This picture does not apply in the
case of the Crab pulsar where pulses are emitted in phase and are
on average polarized along the axis of rotation at all wavelengths
(from radio to $\gamma$-rays; S\l owikowska {\em et al.}~2009). We
can thus argue that, in that particular case, radio emission may
just be the low-energy part of the observed high-energy curvature
radiation emission amplified by coherence effects, as proposed in
C09. It would thus be worthwhile to search for a radio component
in phase and with the same polarization as the high-energy one in
the Fermi pulsars too. If such component is not discovered, it
would be interesting to study why radio coherence from the pulsar
synchrotron is suppressed in most (but not all) pulsars.

\section*{Acknowledgments}
We would like to thank Drs. Alice Harding and Demosthenes Kazanas
for their insightful comments and criticism on an earlier version
of this manuscript.

\appendix
\section[]{Numerical resolution}

\begin{figure*}
\begin{center}
\includegraphics[width=13.2cm]{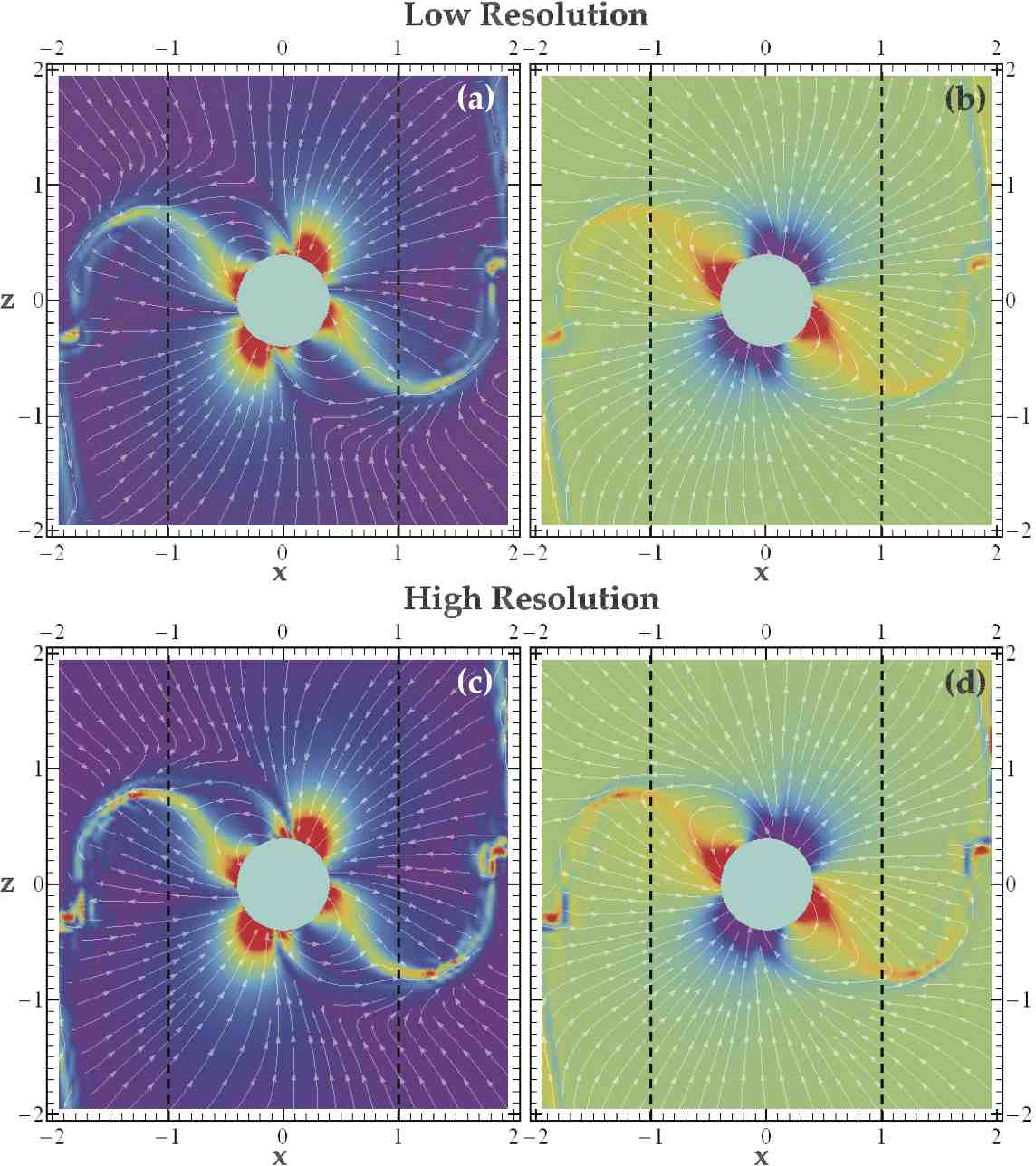}
\end{center}
\caption{Comparison between our standard $\delta=0.04 r_{lc}$ and
a higher $\delta=0.025 r_{lc}$ grid resolutions. Same as Fig.~2
for $\alpha=50^o$ and $t=0$.} \label{resolution}
\end{figure*}

In order to investigate the influence of the numerical grid
resolution $\delta$ on our results, we compared two simulations
for an inclined rotator with $\alpha=50^o$, one with our standard
grid resolution $\delta=0.04 r_{lc}$, and one obtained with higher
resolution $\delta=0.025 r_{lc}$ (Fig.~\ref{resolution}).
In the latter we obtain the same overall magnetic field
distribution, only it is sharper. In particular:
\begin{enumerate}
\item We better resolved random plasmoid formation outside the
light cylinder. \item We obtained a sharper (1-2 computational
cells wide) return current structure along the positively charged
part of the separatrix. This supports our conclusion that the
singular return current sheet discovered in CKF survives also in
the inclined rotator. \item The positively charged spiral feature
outside the light cylinder mentioned in \S~2 is also present. This
supports our view that it is indeed a real feature of the
magnetosphere.
\end{enumerate}

\label{lastpage}
\end{document}